# Extremely High Thermal Conductivity of Graphene: Experimental Study

A. A. Balandin[1,2,*], S. Ghosh[1], W. Bao[3], I. Calizo[1], D. Teweldebrhan[1], F. Miao[3] and C. N. Lau[3]

[1]Nano-Device Laboratory, Department of Electrical Engineering, University of California – Riverside, Riverside, California 92521 USA

[2]Materials Science and Engineering Program, Bourns College of Engineering, University of California – Riverside, Riverside, California 92521 USA

[3]Department of Physics and Astronomy, University of California – Riverside, Riverside, California 92521 USA

**Abstract**

We report on the first measurement of the thermal conductivity of a suspended single layer graphene. The measurements were performed using a non-contact optical technique. The near room-temperature values of the thermal conductivity in the range ~ $(4.84\pm0.44)\times10^3$ to $(5.30\pm0.48)\times10^3$ W/mK were extracted for a single-layer graphene. The extremely high value of the thermal conductivity suggests that graphene can outperform carbon nanotubes in heat conduction.

[*] Corresponding author; E-mail address: balandin@ee.ucr.edu ; web-address: http://ndl.ee.ucr.edu

Graphene – a recently discovered form of carbon[1], which consists of only one plain layer of atoms arranged in a honey-comb lattice – exhibits a number of intriguing properties[1-8]. Its extraordinary high room-temperature (RT) carrier mobility[1,8], conductance quantization[5], possibilities of inducing a band-gap through the lateral quantum confinement[8], and prospects for epitaxial growth[9] make graphene a promising material for future electronic circuits. Despite theoretical suggestions that graphene may also have unusually high thermal conductivity[10-12] no measurements were reported to date to support this claim.

With the continuously decreasing size of electronic devices and increasing dissipation power density in downscaled circuits, one observes a tremendous growth of importance of materials that can conduct heat efficiently. CNTs are known to have very high thermal conductivity[13-14] $K$ with the experimentally determined RT value $K \approx 3000$ W/mK for an individual multi-wall carbon nanotube (MW-CNT)[15] and $K \approx 3500$ W/mK for an individual single-wall carbon nanotube (SW-CNT).[16] The RT thermal conductivity in the range 1750 – 5800 W/mK was reported for the crystalline "ropes" of SW-CNTs.[17] These values exceed those of the best bulk crystalline thermal conductor – diamond – which has the thermal conductivity in the range $K=1000 – 2200$ W/mK.[18] Theoretical calculations of the thermal conductivity of CNTs mostly support the experimental results for individual CNTs, although some discrepancy exists. Molecular dynamics (MD) simulations suggested an unusually high value, $K \approx 6600$ W/mK, for an insolated CNT at RT.[19] Another MD study found RT thermal conductivity in the range 1500 – 3000 W/mK for SW-CNT.[20]

In spite of the importance of the knowledge about the thermal conductivity of graphene no experimental data has been reported to date. The latter can be explained by the fact that the conventional techniques for measuring the thermal conductivity such as thermal bridge, 3ω method or "laser-flash", are not well suited for single-layer graphene (SLG). The 3ω method is good for measuring the cross-plane thermal conductivity, and requires substantial temperature drop over the thickness of the examined film.[21-22] Graphene with the thickness of one atomic layer and expected high thermal conductivity cannot satisfy such a requirement. The direct thermal-bridge measurements of graphene are possible in principle but very challenging technologically.

Here we undertook an unconventional approach for the non-contact measurement of the thermal conductivity of graphene by using the confocal micro-Raman spectroscopy. Several factors, specific for graphene, made it possible. Graphene has clear signatures in Raman spectra[23-25]. We have also recently discovered that the *G* peak in graphene spectra manifest a strong temperature dependence.[26] The *G* peak's temperature sensitivity allows one to monitor the local temperature change produced by the variation of the laser excitation power focused on a graphene layer. In a properly designed experiment, the local temperature rise as a function of the laser power can be utilized to extract the value of the thermal conductivity. A Raman spectroscopy – based technique has been previously used successfully for measuring the thermal conductivity of poorly heat conducting materials and thin films.[27-28] At the same time, there are major differences in heat spreading in graphene from that in conventional materials. For this reason we had to develop a new measurement methodology and derive expressions suitable for the thermal conductivity extraction for graphene. The Raman spectroscopy-based measurement of the thermal conductivity is not suitable for the bulk crystalline materials with the high thermal conductivity because of the rapid escape of heat, produced by the laser excitation, in three-dimensional systems. The latter prevents a local temperature rise detectable with the Raman spectroscopy for reasonable excitation power levels. Luckily, graphene has a thickness of only one atomic layer. Thus, if we suspend graphene over a trench and heat in the middle, the heat is forced to propagate in-plane through the layer with the thickness[24] $h= 0.35\pm0.01$ nm toward the heat sink. The extremely small cross-section area of the heat conduction channel makes the detection of the local temperature rise possible.

Graphene samples have been obtained by the mechanical cleavage of bulk graphite using the standard technique[1]. The SLG flakes connected to multi-layer graphenes were selected using the Raman spectroscopy and *2D*-band deconvolution[23,25]. Figure 1 shows that the suspended graphene flake has the Stokes *G* peak at 1583 cm$^{-1}$ and a symmetric *2D* band around 2700 cm$^{-1}$, which is consistent with the reported SLG spectra[23-26]. We fabricated trenches on a number of Si/SiO$_2$ substrates by the reactive ion etching and placed long graphene flakes over these trenches to obtain suspended graphene. The nominal depth of the trenches was 300 nm. The width of the trenches varied from 2 to 5 µm. One can see in this figure 5-µm-wide SLG and 1-µm-wide few-layer graphene (FLG) bridging the two side of the trench. The scanning electron microscopy (SEM) image in Figure 2 gives a close-up view of the suspended graphene layers over the 3-µm-wide trench.

The schematic of the experiment is shown in Figure 3. The measurements were carried out using the confocal micro-Raman spectroscopy[25-26], which allowed us to restrict the graphene sampling volume to the suspended portion of the flake. The laser light was focused in the middle of the suspended SLG with the spot size of about 0.5-1.0 μm. Since the thermal conductivity of air is negligible (~0.025 W/mK) the heat generated in SLG due to the laser excitation has to escape propagating laterally through the extremely thin graphene layer. As a result, even a small power dissipated in the middle of SLG can lead to a detectable rise of the local temperature. Many of the examined grapheme flakes were connected at the periphery to bulk graphite. The attached large graphitic peaces at the distance of 9-10 μm from the trench edge acted as heat sinks. The heat sink temperature does not change during the experiment for the low power levels involved. The thermal coupling of graphene to Si substrate is also small due to the microscopic corrugations of the partially suspended graphene, low thermal conductivity of the oxide (~1 W/mK) and very large thermal interface resistance. As a result the heat wave generated over the suspended portion of graphene continues to propagate all the way to the heat sink. From the available excitation wavelengths we have chosen 488-nm laser light for this experiment. The shorter excitation wavelength in the ultraviolet range, e.g. 325 nm, is strongly absorbed and good for heating the sample surface but it does not provide clear Raman signatures for graphene. The longer wavelength, e.g. 633 nm, excites informative scattering spectra from graphene but does not produce local heating as efficiently as 488 nm laser light.

The heat conduction through the surface with the cross-sectional area $S$ can be evaluated as $\partial Q/\partial t = -K \oint \nabla T \cdot dS$, where $Q$ is the amount of heat transferred over the time $t$ and $T$ is the absolute temperature. The phonon mean free path data reported for CNTs[14-16] suggest that the thermal transport in graphene over μm-size flakes has to be at least partially diffusive. The exact shape of the heat front propagating through SLG is not known and depends on the shape of the flake and its edges. Two limiting cases can be considered: the radial heat flow from the middle of the suspended flake toward its borders, and a plane-wave front in two opposite directions toward the trench edges. The first case is appropriate when the laser-induced hot-spot is much smaller than the suspended graphene layer size while the second case corresponds to the situation when the laser hot-spot is comparable to the layer width $W$. Writing the uniform radial heat flow equation for the two laser excitation power levels $P_1$ and $P_2$, which correspond to the two hot spot temperatures, we obtained the expression for the thermal conductivity $K=(1/2\pi h)(\Delta P/\Delta T)$, where $h$ is SLG thickness and the local temperature rise $\Delta T$ is due to the changing heating power $\Delta P = P_2 - P_1$.

Since the excitation power levels are relatively low the *G* peak position linearly depends on the sample temperature[26] $\omega = \omega_o + \chi_G T$. The final expression for the thermal conductivity in the radial case can be written as $K = \chi_G (1/2h\pi)(\delta\omega/\delta P)^{-1}$, where $\delta\omega$ is a small shift in the *G* peak position due to the variation $\delta P$ in the heating power on the sample surface. Analogous considerations for the case of the plane-wave heat front lead to the expression $K=(L/2S)(\Delta P/\Delta T)$, where $L$ is the distance from the middle of the suspended SLG to the heat sink with the ambient $T$ and $S = h \times W$. Finally, the thermal conductivity can be evaluated as $K = \chi_G (L/2hW)(\delta\omega/\delta P)^{-1}$. The above equations include the change in the heating power $\delta P$ on graphene rather than its absolute value or the total power change. A detail analysis of the method accuracy with the estimate of the power loss in the trench bottom is given at the end of this letter.

We measured the excitation power dependence of the Raman *G* peak for a number of the long graphene flakes suspended over the trench and connected to the large graphite heat sinks. The examined samples were described better by the equation for the heat plane waves propagating in two opposite directions toward heat sinks. The excitation power at the graphene sample location was determined with the power meter. The increase in the excitation power led to the increase in the intensity count and red shift of the *G* mode peak. The intensity increase is clearly seen in Figure 4. The red shift indicates a rise in the local temperature in the middle of the suspended graphene. Figure 5 shows the *G* peak position dependence on the total dissipated power $P_D$ for a high-quality suspended SLG. In this figure the peak position change is referenced to the value, which corresponds to the lowest excitation power. The extracted slope is $\delta\omega/\delta P_D \approx -1.29$ cm$^{-1}$/mW. The temperature coefficient $\chi_G$ for the *G* peak of graphene was determined by us previously for SLG produced by the same technique[26]. It was accomplished by keeping the low-level of laser excitation constant and externally changing the temperature of SLG placed in the cold-hot cell. Extracting from the measured $\delta\omega/\delta P_D$ the value for graphene $\delta\omega/\delta P$ though the procedure outlined below, and substituting together with $\chi_G = -1.6 \times 10^{-2}$ cm$^{-1}$/K and $h$ to the second equation for the thermal conductivity we obtain for the set of SLG samples the averaged values in the range $(4.84\pm0.44)\times10^3$ to $(5.30\pm0.48)\times10^3$ W/mK. The extracted thermal conductivity is on the high-end of the values reported for CNTs or exceeds them.

The thermal conductivity of SLG with the RT electrical resistance of 1-4 kΩ is mostly due to the acoustic phonons since the electron contribution to *K*, estimated from the Wiedemann-Franz law, is negligible. This is consistent with the observation for the thermal conductivity of individual CNTs[15,17]. It is illustrative to compare the obtained result with the Klemens[29] calculation of the thermal conductivity of graphite in the basal plane (*a*-plane). In his work, the thermal conductivity of the highly oriented pyrolytic graphite (HOPG) was treated in terms of the two-dimensional phonon gas. For the intrinsic, i.e. Umklapp scattering limited thermal conductivity, Klemens found *K*=1910 W/mK at RT. Our experimental value for graphene is approximately a factor of three larger. Better thermal conduction properties of graphene can be related to the smaller Grüneisen parameter $\gamma$ for the pertinent phonon modes as predicted from the first-principle calculations[30] since $K \sim 1/\gamma^2$. A recent MD study of the in-plane thermal conductivity of graphite[31] reported *K*=1000 W/mK which is approximately two times below the Klemen's result. The same study found *K*=2980 W/mK for CNTs and revealed a strong dependence of the thermal conductivity on the vacancy and defect concentration[31]. The situation is likely similar for graphene.

The extremely high value of the phonon thermal conductivity of the strictly two-dimensional material system such as graphene comes in sharp contrast with the reduced phonon thermal conductivity (as compared to their bulk values) of the quasi-two-dimensional systems such as semiconductor quantum wells (thin films)[32] and quasi-one-dimensional systems such as quantum wires (nanowires)[33]. The reduction of the phonon (lattice) thermal conductivity in quantum wells and nanowires comes as a result of the phonon – rough boundary scattering or phonon spatial confinement effects[32-34]. Another interesting observation is a huge range of the thermal conductivities of carbon materials, which spans from some of the lowest RT values of ~0.2 W/mK reported for diamond-like carbons[35] to up to ~5300 W/mK measured in this work for graphene. In this sense, carbon materials can serve both as thermal insulators (some of the diamond-like carbons or amorphous carbons) as well as the "heat superconductors" (graphene).

Let us now discuss in more details the data extraction procedure. Before measuring the thermal conductivity we estimated the effect produced by the light, which penetrates through the suspended graphene. Although the light is focused on graphene surface, it penetrates down to Si substrate and generates Stokes power. The power measured with the detector can be split into two terms $P_D=P+P_{Si}$, where $P$ is the power to be dissipated in graphene and $P_{Si}$ is the power to be lost in Si substrate. Note that since we measure the power at the sample position and not at the laser output the losses in the spectrometer do not affect the accuracy of our method. The amount of the power $P$ absorbed by the suspended graphene can be evaluated through the calibration procedure with HOPG, which was used to produce the graphene samples, via the analyses of the integrated *G* peak intensities from graphene and HOPG.

The power absorbed in graphene is given by $P = I_o A(1-\exp(-\alpha_G a_G)) \approx \alpha_G a_G I_o A,$ where $A$ is the illuminated area, $I_o$ is the laser intensity on the surface, $\alpha_G$ is the absorption coefficient and $a_G$ is the monolayer thickness. The integrated Raman intensity from SLG can be written as[36] $\overline{\Delta I_G} = N\sigma_G I_o,$ where $N/A$ is the surface number density of the scattering atoms and $\sigma_G$ is scattering cross section. Thus, the Raman intensity can be related to the absorbed power as $\overline{\Delta I_G} = (N/A)(\sigma_G/\alpha_G a_G)P$. When the same laser beam is focused on the calibration HOPG sample, the measured power $P_D \approx I_o A$. The scattered intensity from HOPG can be obtained by summation over all *n* layers $\overline{\Delta I_{HOPG}} = N\sigma_H I_o \sum_{n=1}^{\infty} \exp(-2\alpha_H a_H n) \approx N\sigma_H I_o (\exp(2\alpha_H a_H)-1)^{-1}$. The later reduces to $\overline{\Delta I_{HOPG}} \approx (1/2)(N/A)(\sigma_H/\alpha_H a_H)P_D.$ Here $\alpha_H$ and $\sigma_G$ are the absorption coefficient and scattering cross-section for HOPG with the monolayer of thickness $a_H$.

Defining the ratio of the integrated intensities as $\varsigma = \overline{\Delta I_G}/\overline{\Delta I_{HOPG}}$, we can express the power absorbed in graphene through the power measured by the detector as $P = (\varsigma/2)[\sigma_H \alpha_G a_G / \sigma_G \alpha_H a_H]P_D.$ The term in the square brackets is very close unity since it consists of the ratios of the in-plane microscopic material parameters for essentially the same material. The value of $\zeta$, determined experimentally, completes the calibration. For a number of examined samples, we found $\zeta$ to be 0.26. The distribution of power between the suspended graphene and substrate may depend on the conditions of the experiment and was checked for each experimental run. The presence of the microscopic corrugations in the partially suspended graphene has been recently reported in Ref. [37]. The latter supports our assumption about relatively weak coupling to the substrate. We also monitored the positions of $W_2$ and $W_3$ Si-O-Si stretching bonds [38] in the range from 800 to 1100 cm$^{-1}$ at different power levels to confirm that SiO$_2$ layer is not strongly heated during the measurement. That was possible because despite focusing on the suspended portion of graphene, the smallest spot size is ~0.5-1.0 μm [39], and some of the excitation light covers SiO$_2$ layer under graphene near the trench edges. Finishing the measurement error analysis we estimated that the inclusion of the finite reflectivity of HOPG and Si trench bottom introduces minor change to the final result (within 9% of the experimental uncertainty).

In conclusion, we reported the first experimental study of the thermal conductivity of single-layer graphene. Using a non-contact optical based technique we discovered that graphene manifests an extraordinary high RT thermal conductivity of up to $(5.30\pm0.48)\times10^3$ W/mK. The measurements were performed for an individual single-layer graphene suspended over a wide trench in Si/SiO$_2$ substrate. The extracted thermal conductivity of graphene is larger than any experimental values reported for individual suspended CNTs and corresponds to the upper bound of the highest values reported for single-wall CNT bundles. The superb thermal conduction property of graphene is beneficial for the proposed electronic applications and establishes graphene as an excellent material for thermal management.

*Acknowledgements*

A.A.B. acknowledges the support from DARPA – SRC through the FCRP Center on Functional Engineered Nano Architectonics (FENA). A.A.B. and C.N.L acknowledge the support from DARPA – DMEA through the UCR – UCLA – UCSB Center for Nanoscience Innovations for Defense (CNID). A.A.B. is indebted to Drs. E.P. Pokatilov and D.L. Nika (Moldova State University) for checking the derivations; Drs. M. Makeev (NASA Ames), N. Kalugin (New Mexico Tech), M. Kuball (University of Bristol) and C. Dames (UC Riverside) for critical reading of the original version of the manuscript; and to Dr. A.C. Ferrari (Cambridge University) for insightful discussions on graphene.

**FIGURE CAPTIONS**

**Figure 1:** Raman spectrum of graphene showing the *G* peak and *2D* band features characteristic for single-layer graphene.

**Figure 2:** Scanning electron microscopy image of the suspended single and few-layer graphene across a trench in $SiO_2$/Si wafer.

**Figure 3:** Schematic of the experiment showing the excitation laser light focused on a graphene layer suspended across a trench. The focused laser light creates a local hot spot and generates a heat wave inside graphene propagating toward heat sinks.

**Figure 4:** *G*-peak region of the Raman spectrum from a single-layer graphene recorded at two excitation power levels.

**Figure 5:** The shift in *G*-peak spectral position vs. change in total dissipated power. The spectra are excited at 488 nm and recorded at room temperature in the backscattering configuration.

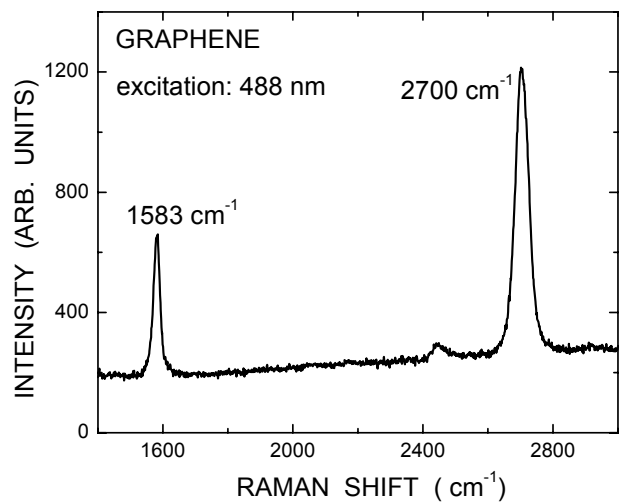

Figure 1 of 5

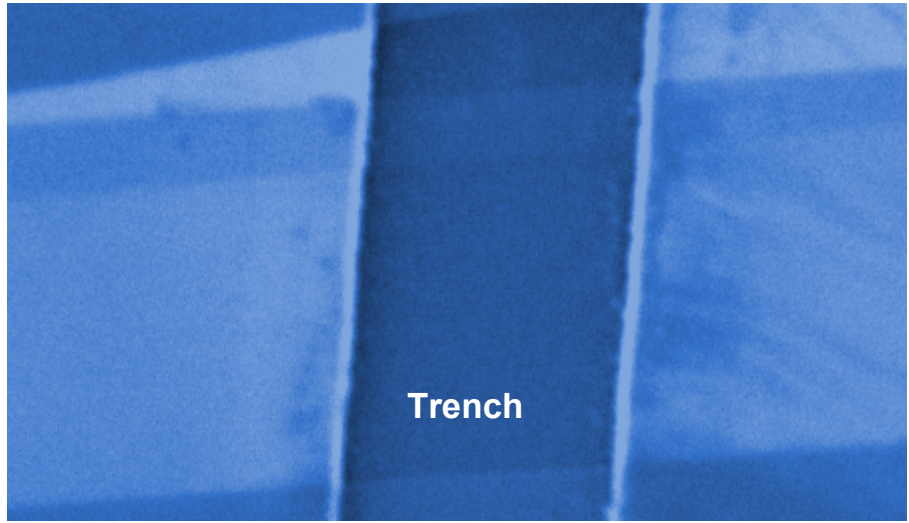

Figure 2 of 5

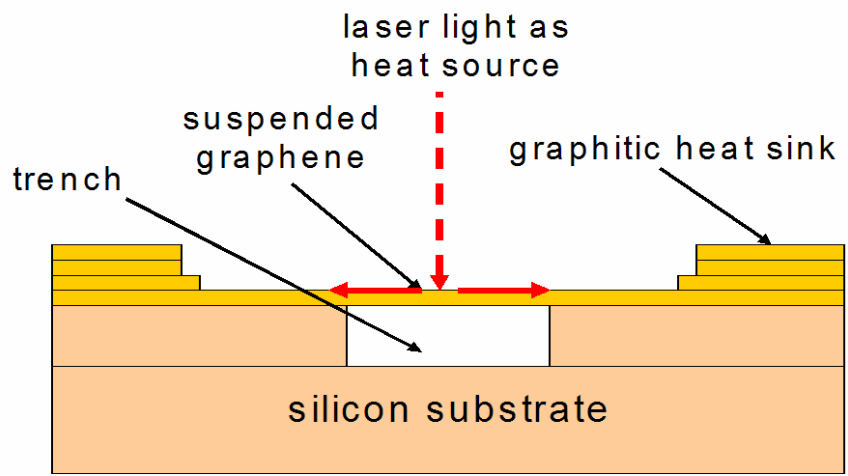

Figure 3 of 5

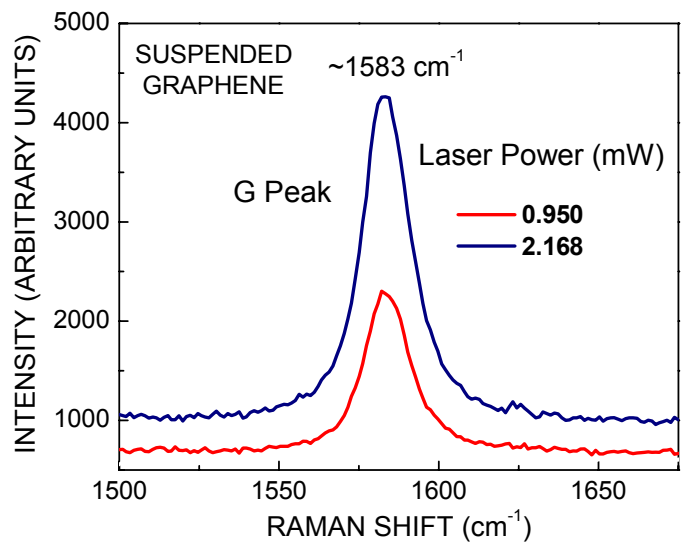

Figure 4 of 45

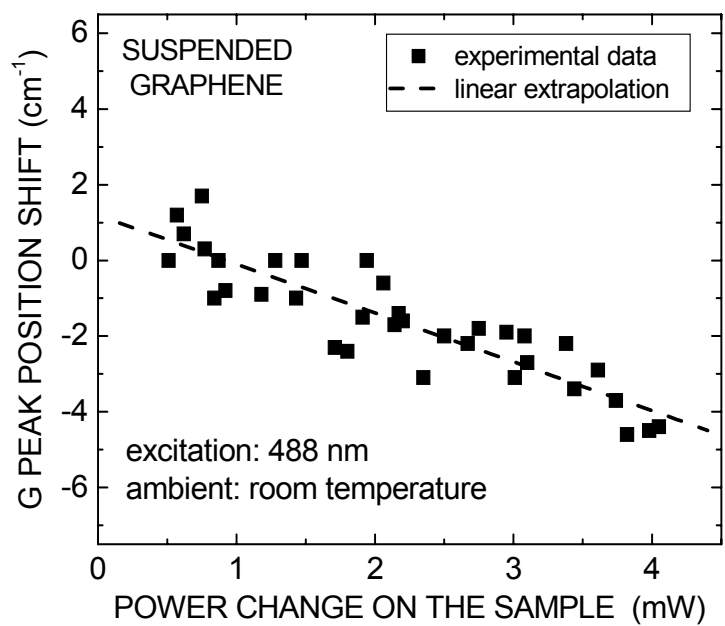

Figure 5 of 5